\documentclass[conference, 9pt]{IEEEtran}
\IEEEoverridecommandlockouts
\usepackage{cite}
\usepackage{amsmath,amssymb,amsfonts}
\usepackage{algorithmic}
\usepackage{graphicx}
\usepackage{textcomp}
\usepackage{xcolor}
\usepackage{hhline}
\usepackage{hyperref}
\usepackage{multirow}
\usepackage{makecell}
\def\BibTeX{{\rm B\kern-.05em{\sc i\kern-.025em b}\kern-.08em
    T\kern-.1667em\lower.7ex\hbox{E}\kern-.125emX}}

\newif\iftodo
\todotrue

\newif\ifneedconfirmation
\needconfirmationtrue

\begin{document}


\title{Generative Data Augmentation Challenge: Zero-Shot Speech Synthesis for Personalized Speech Enhancement\thanks{This material is in part based on work supported by the National Science Foundation under grant numbers, 2512987 and 1910940.

The authors appreciate Apoorv Vyas at Meta for his help in establishing the virtual speaker dataset.}}
\author{\IEEEauthorblockN{Jae-Sung Bae$^{1,7}$, Anastasia Kuznetsova$^2$, Dinesh Manocha$^3$, John Hershey$^4$, Trausti Kristjansson$^{5,6}$, and Minje Kim$^{1,5,8}$}
\IEEEauthorblockA{
$^1$\textit{University of Illinois Urbana-Champaign}, $^2$\textit{Indiana University}, $^3$\textit{University of Maryland},\\$^4$\textit{Google Research}, $^5$\textit{Amazon Lab126}, $^6$\textit{Reykjavik University}\\
$^7$jb82@illinois.edu, $^8$minje@illinois.edu}
}

\maketitle

\begin{abstract}
This paper presents a new challenge that calls for zero-shot text-to-speech (TTS) systems to augment speech data for the downstream task, personalized speech enhancement (PSE), as part of the Generative Data Augmentation workshop at ICASSP 2025. Collecting high-quality personalized data is challenging due to privacy concerns and technical difficulties in recording audio from the test scene. To address these issues, synthetic data generation using generative models has gained significant attention. In this challenge, participants are tasked first with building zero-shot TTS systems to augment personalized data. Subsequently, PSE systems are asked to be trained with this augmented personalized dataset. Through this challenge, we aim to investigate how the quality of augmented data generated by zero-shot TTS models affects PSE model performance. We also provide baseline experiments using open-source zero-shot TTS models to encourage participation and benchmark advancements. Our baseline code implementation and checkpoints are available online\footnote{\href{https://sites.google.com/view/genda2025}{https://sites.google.com/view/genda2025}}.
\end{abstract}


\section{Introduction}

In this paper, we introduce a new challenge that accompanies the Generative Data Augmentation workshop at ICASSP 2025. The main research problem the challenge addresses is the typical data shortage issues when a machine-learning model is developed for a particular target user so that the model respects and exploits individuals' specificity. This type of problem contrasts the remarkable advancements driven by deep neural networks (DNNs) in speech technology, including automatic speech recognition (ASR) \cite{whisper}, text-to-speech (TTS) \cite{vits, yourtts, xtts, valle}, and speech enhancement (SE) \cite{se_1, se_2}. These models are typically large in size and trained on extensive datasets, designed to perform robustly across diverse inputs; we refer to such systems as \textit{generalist} systems. However, due to their large size, these systems are challenging to run directly on user devices, necessitating the transfer of user data to servers, which can raise privacy concerns. To address this, smaller \textit{personalized} models have recently been proposed \cite{apple_pasr, minje_pse_2}. These models, being compact, can operate on user devices and achieve high performance by focusing on individual users. In theory, personalized models can be created by fine-tuning pre-trained models using individual-specific data. However, gathering enough personalized data for each individual remains a significant challenge due to privacy concerns and technical difficulties in recording clean voices at test time.

Data augmentation is an essential technique to improve the performance of many DNN models. In image processing, a variety of data augmentation techniques are employed, ranging from simple methods like rotating or flipping images \cite{image_dataaug_survey} to generating new images using other state-of-the-art image generation models \cite{image_da_diffusion}. Similarly, numerous data augmentation techniques are also being explored in the field of speech \cite{spechaug, apple_pasr, minje_pse_2, latent_filling}. SpecAugment \cite{spechaug} is one of the most commonly used data augmentation methods in speech self-supervised learning and automatic speech recognition models. Recently, with advancements in TTS models enabling near-human-level speech generation, approaches using TTS for data augmentation have been actively attempted \cite{apple_pasr, minje_pse_2, tts_for_DA, tts_keyword_da}. 

\begin{figure}[t]
\centering
{
    \centering
    \begin{minipage}[t]{\linewidth}
      \centering
      \centerline{\includegraphics[width=0.72\linewidth]{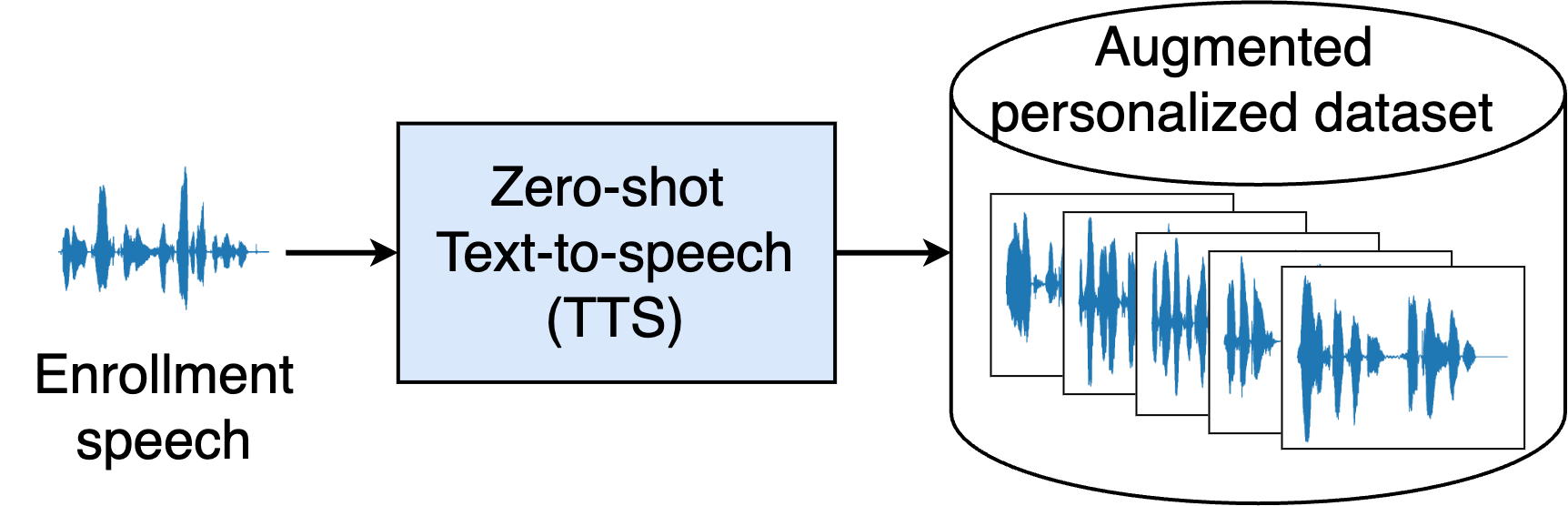}}
      \centerline{(a)}
    \end{minipage}
    \centering
    \begin{minipage}[t]{\linewidth}
      \centering
      \centerline{\includegraphics[width=0.82\linewidth]{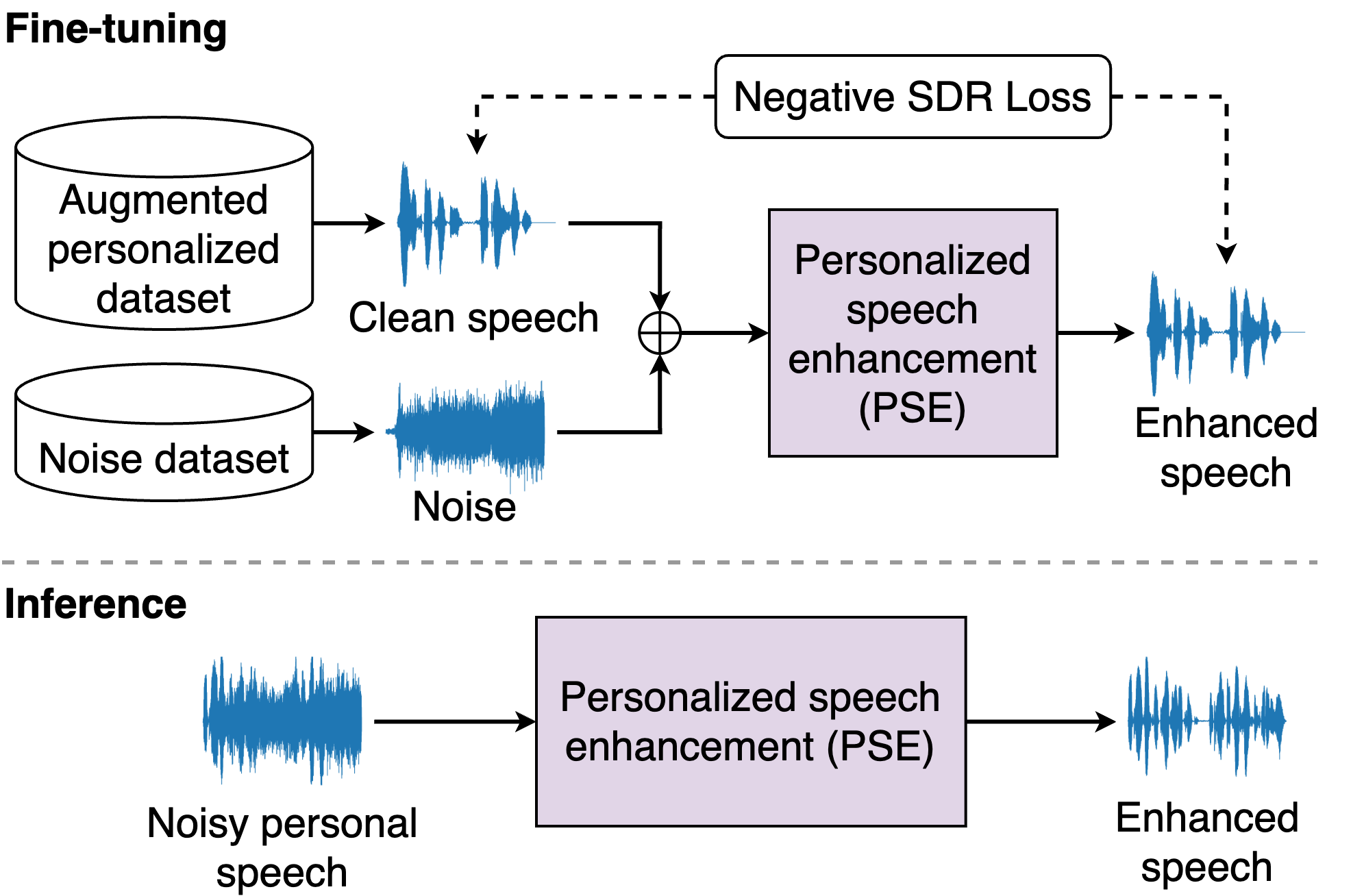}}
      \centerline{(b)}
    \end{minipage}
    \vspace{-0.1in}
    \caption{Overall flow of our challenge. (a) First, participants augment the personalized data using a zero-shot TTS model. (b) Next, organizers ask participants to fine-tune the PSE model with this augmented personalized data, followed by an inference phase in which enhanced speech is generated for evaluation.}
    \label{fig}
}
\end{figure}

Furthermore, the performance of zero-shot TTS systems \cite{yourtts, xtts, valle}, which can generate speech that mimics a target speaker’s characteristics from just a short speech signal, has significantly improved. These systems, with the ability to generate an unlimited number of speech samples that reflect a target speaker’s characteristics from a single speech signal, are gaining attention as a data augmentation approach to address the data shortage problem of personalized models \cite{minje_pse_2, ZSTTS_DA_for_ASR}. Although prior research has demonstrated this potential, studies exploring the relationship between various zero-shot TTS models’ performance and downstream tasks remain limited.

To conduct a more comprehensive investigation into how zero-shot TTS systems can benefit downstream personalized systems, we propose the \textbf{Zero-shot Speech Synthesis Challenge for Personalized Speech Enhancement (PSE)}. We hypothesize that the higher the quality of augmented speech samples generated by zero-shot TTS systems, the better the performance of downstream tasks fine-tuned using those samples. However, we also believe different performance evaluation factors of augmented speech, such as the precise perceptual similarity between synthesized and original speakers, perceived naturalness of the speech, and the quality of speech signals may have different implications in different downstream tasks. With the proposed challenge, as illustrated in Fig. \ref{fig}, we aim to highlight the potential difference between the method for evaluating TTS results and their usefulness in the particular downstream task, personalized speech enhancement. In the first phase, the organizers provide a set of enrollment speech samples for test speakers, and participants are tasked with generating speech samples with target speaker characteristics using their zero-shot TTS systems. The generated speech is then evaluated based on speaker similarity, perceptual quality, and intelligibility. In the second phase, participants are asked to train PSE systems using the augmented personalized speech dataset, with performance evaluated on enhanced speech quality. 

We conduct baseline experiments for this challenge, providing a detailed description of the experiments and results in this paper. We open-source our baseline model, offering participants a reference for their experiments.

\begin{table*}[ht]  
\caption{Description of the datasets used in our challenge and baseline experiments}
\begin{center}
\resizebox{.9\textwidth}{!}{  \begin{tabular}{|c|c|l|c|}
\hline
\multicolumn{1}{|c|}{\textbf{Duration}} & \textbf{Quantity} & \multicolumn{1}{c|}{\textbf{Descriptoin}} & \textbf{Corpus}\\
\hline
1 utt/spkrs & \multirow{4}{*}{20 spkrs} & Enrollment speech of target speakers for zero-shot TTS  system. & \multirow{4}{*}{\makecell[l]{10 spkrs from LibriTTS \cite{libritts} test-\\clean and another 10 spkrs from the \\virtual speaker dataset (Sec. \ref{sec:data:virtual})}} \\ 
\cline{1-1} \cline{3-3}
9 utts/spkrs &  & Test speech samples for PSE system evaluation. & \\ 
\cline{1-1} \cline{3-3}
50 utts/spkrs &  & \makecell[l]{Test speech samples for zero-shot TTS system evaluation. Forty utterances\\ were used to train the \textbf{GT-6min} model in the baseline experiments.}& \\
\hline
5 noises/spkrs & 88 noises & \makecell[l]{Injection noise used during PSE training; a unique set of 5 noise sources\\are used for each speaker.} & MUSAN \cite{musan} \\ \hline
40 utts/sprks & \multirow{3}{*}{20 spkrs} & \makecell[l]{Synthesized speech from zero-shot TTS models, used to train \textbf{6min} \\ models in the baseline experiments.} & \multirow{3}{*}{\makecell[l]{Three augmented personalized datasets\\from three baseline zero-shot TTS \\systems, respectively.}}\\ \cline{1-1} \cline{3-3}
180 utts/sprks &  & \makecell[l]{Additional synthesized speech from zero-shot TTS models, used to train \\ \textbf{30min} models in the baseline experiments.} & \\
\hline
\end{tabular}}
\vspace{-0.15in}
\label{tab:data}
\end{center}
\end{table*}

\section{Challenge Description}
There are two main technical components in this challenge: the generative models that the participants develop to synthesize personalized speech utterances (i.e., the zero-shot TTS systems) and the downstream PSE task where the synthesized personal speech signals are used as the training target of the PSE systems. The challenge organizers will evaluate the submissions mainly based on their usefulness on the PSE tasks, while some basic speech quality estimation will also be conducted for a comprehensive evaluation of the TTS systems.

\subsection{Zero-Shot TTS Models}
Participants are required to develop a zero-shot TTS system, which is supposed to synthesize new utterances from a short enrollment signal of the target speaker. The organizers will provide from 3 to 14 second-long enrollment signal per target speaker, who is one of ten randomly chosen speakers from the LibriTTS test-clean dataset \cite{libritts} (Sec. \ref{sec:data:real}) or another ten virtual speakers we create for the challenge (Sec. \ref{sec:data:virtual}). Additionally, the organizers provide 50 text sentences per target speaker for the participating TTS systems to synthesize corresponding speech signals. This will be used to evaluate the overall quality of generated utterances from zero-shot TTS systems. For the evaluation details of the zero-shot TTS system, please refer to Section \ref{sec:eval:tts}.  

Synthesized signals from zero-shot TTS systems are supposed to be used to train the PSE systems, whose SE performance is assumed to be associated with the TTS systems' performance, i.e., the better the synthesized speech preserves the target speaker identity and speech quality, the more useful it is to personalize a speech enhancement system. The organizers impose no restrictions on the number of utterances the zero-shot TTS system generates for training the PSE systems. While there are no other specific restrictions on the zero-shot TTS model in terms of its architecture or how it is trained, \textbf{the LibriTTS test-clean dataset must be excluded in the TTS model's training as it is used for testing.}

\subsection{The Downstream Personalized Speech Enhancement Task}\label{sec:pse}

PSE is a category of SE methods designed to train specialized models for individual users at test time. Here, we assume that PSE systems are personalized not only to specific speakers but also to the noise environments that particular users frequently encounter. Unlike the speaker and noise-agnostic SE models, which are trained to generalize to any arbitrary test speakers and noises, PSE models focus on the specific speaker and noise types of interest. Ideally, personalization could be achieved by training an SE model by setting up the clean speech of the target speaker as the denoising target and noisy speech with specific noise types as the noisy input. However, in practice, such personal data are difficult to acquire due to privacy concerns and the technical difficulties of recording the clean voice samples from the test-time user. 

Successful TTS systems can resolve the data shortage issue. They can synthesize virtually as many clean speech signals as a PSE model needs for its personalized training. Since the main goal of personalization is to narrow down its usage to the target user, it is also crucial for the downstream task to be able to use personality-preserving, high-quality speech signals. Likewise, PSE performance is sensitive to the quality of participants' TTS systems, making it a suitable task for evaluating the generative data augmentation systems. 

Participants are asked to use their synthesized speech to develop 20 PSE models for 20 target speakers, respectively. In addition, the organizers provide a test set consisting of 45 noisy utterances for each speaker, which are mixtures of nine clean utterances and five speaker-specific types of noise sources (Sec. \ref{sec:data:noise}). The input signal-to-noise ratio (SNR) is randomly selected from $\{-2.5, 0, 2.5\}$ dB. Participants are required to submit enhanced speech for these samples. 
For a fair comparison, participants are first requested to enhance the noisy signals with the baseline PSE model architecture that organizers provide. Then, participants can optionally provide results from their own model architectures. We strongly encourage participants to build lightweight PSE models as one of the main benefits of personalization is about being able to reduce model size. To this end, the complexity of the model must be reported as well.

\section{Challenge Datasets}
\label{sec:data}

\subsection{Real-World Speakers}\label{sec:data:real}
From the LibriTTS \cite{libritts} dataset's test-clean fold, five male and five female speakers are randomly selected as the personalization target. After excluding utterances that are either shorter than 3 seconds or longer than 16 seconds, we collect 60 utterances from each speaker that are then divided into three subsets: one utterance for enrollment, 50 for evaluating the generated speech directly, and nine for testing the PSE systems.

\subsection{Virtual Speakers}\label{sec:data:virtual}
In order to propose a solution and draw attention to the potential ethics and privacy issues that revolve around using the speech corpus as a seed to synthesize personalized speech, we additionally adopted ten virtual speakers as target speakers. Similar to the real-world speakers, these virtual speakers include five male and five female speakers. However, unlike real-world speakers, who primarily exhibit reading-style speech, we aimed to leverage the diversity of TTS models to generate virtual speakers with various accents and emotional expressions. Note that all the challenge configurations remain the same for these ten speakers as well. The significant difference is that these virtual speakers were generated using {a state-of-the-art TTS system provided by Meta}.

\subsection{Noise Sources}\label{sec:data:noise}
Not only to personalize the PSE systems for speakers but also to adapt to noisy environments, we designate five specific noise types per target speaker, which are randomly chosen from the \textit{sound-bible} subset of the MUSAN \cite{musan} dataset.

\section{Evaluation Metrics}\label{sec:eval}

\subsection{Basic TTS Performance Evaluation}\label{sec:eval:tts}
Although the challenge's main objective is to prove the generative models' capabilities in the downstream task, we can conduct some basic performance evaluation of the synthesized data themselves. 
To this end, the zero-shot TTS system’s performance is evaluated across three aspects: speaker similarity, intelligibility, and perceptual quality. To measure speaker similarity between the generated speech and the reference speech, we calculate the cosine distance of speaker embeddings (SECS). X-vectors are extracted using a pre-trained speaker verification model from SpeechBrain \cite{speechbrain} and utilized to calculate SECS. To assess the intelligibility of the generated speech, we employ the open-source Whisper model \cite{whisper} for speech recognition and calculate the word error rate (WER).

Subjective evaluations are commonly conducted to assess the perceptual quality of TTS models. However, conducting these evaluations across numerous models poses practical challenges. Recently, neural network-based perceptual quality metrics have been widely studied \cite{mosnet, mosnet2} and increasingly adopted \cite{minje_pse_2, svc_utmos1, naturalspeech3_utmos2}. In this work, we chose to use the UTMOS \cite{utmosv2} metric---one of the best-performed MOS prediction networks in VoiceMOS Challenge 2024 \cite{voicemos_challenge}-to evaluate the perceptual quality and naturalness of the generated speech. 

\subsection{Evaluation metrics for PSE models}\label{sec:eval:pse}
To evaluate the performance of the PSE models and examine the effectiveness of incorporating augmented speech from zero-shot TTS systems, we employ four metrics commonly used in SE: signal-to-distortion ratio improvement (SDRI), signal-to-distortion ratio (SDR) \cite{sdr}, extended short-time objective intelligibility (eSTOI) \cite{estoi}, and perceptual evaluation of speech quality (PESQ) \cite{pesq}. All the metrics are based on the direct comparison between the participants' submissions, i.e., the enhanced versions of the noisy test signals and the held-out ground-truth clean speech signals. SDRI measures the improvement in SDR, indicating how effectively the model reduces distortion relative to the input. SDR evaluates the absolute quality of the enhanced signal by comparing the energy ratio between the target and distortion signals, reflecting the fidelity preservation of the model’s output. eSTOI, a measure of intelligibility, assesses how well the model retains short-time temporal patterns. Lastly, PESQ evaluates the perceptual quality of generated speech by simulating human auditory perception. 

\section{Baseline Models}
In this section, we introduce the baseline zero-shot TTS and PSE models and report their performances as a reference for the participants. 

\subsection{Baseline zero-shot TTS models}
We use three open-source zero-shot TTS models as the baseline models: YourTTS \cite{yourtts}, SpeechT5-based zero-shot TTS model \cite{speecht5}, and XTTS \cite{xtts}. 
YourTTS is built on VITS \cite{vits} and conditions it via a speaker embedding extracted from an external pre-trained speaker verification model. SpeechT5 is a pre-trained encoder-decoder model for various spoken language processing tasks, including its TTS application, which we use as a baseline. Finally, XTTS is a zero-shot TTS model built on Tortoise \cite{tortoise}, which incorporates a decoder-only Transformer with some modifications to improve voice cloning and enable faster training and inference.

\subsection{Baseline PSE models}
We adopt the ConvTasNet \cite{convtasnet}-based architecture for our PSE models based on the original PSE model architecture proposed in \cite{minje_pse_1, minje_pse_2}. Following their recipe, the generalist models are first trained on LibriSpeech \cite{librispeech} and FSD50K \cite{fsdk50}, which have clean and noise-mixed speech samples, respectively. To introduce artificial noise addition, we utilize MUSAN \cite{musan} datasets. Given that reduced size is one of the main advantages of the personalized system, we focus on the medium, small, and tiny models from  \cite{minje_pse_1, minje_pse_2}, containing 437K, 224K, and 138.8K parameters, respectively.

Then, we fine-tune the generalist model into a test speaker-specific version for each test speaker, using the personalized speech datasets synthesized by the TTS models. To this end, we run the zero-shot TTS models to generate 40 new clean utterances per speaker. These synthesized signals work as the training target and are mixed with speaker-specific noise sources to create corresponding noisy input mixtures. In addition, we also provide an oracle performance by fine-tuning the generalist model with the ground-truth clean speech, i.e., 40 actual utterances from the same test speaker. Since the duration of these 40 utterances is about six minutes, we refer to these PSE models as \textbf{6min} models. To investigate the performance gains from using additional synthesized speech, we generate 180 additional utterances for fine-tuning, resulting in 220 training utterances (about 30 minutes). We refer to the PSE models with this extended dataset as \textbf{30min} models. Note that fine-tuning uses 10 and 30 validation utterances per speaker for the \textbf{6min} and \textbf{30min} models, respectively. 

We use the negative SDR loss function as in \cite{minje_pse_2}. For the optimization, Adam \cite{adam} is used with a low learning rate of $10^{-6}$. The batch size is 8. We stop fine-tuning if the validation loss does not improve after 20 epochs. The input mixtures are with a randomly selected SNR value from the range of $[-5, 5]$.

\begin{table}[t]  
\caption{Zero-shot TTS results for real-world speakers with 95\% confidence intervals (CIs).}
\begin{center}
\resizebox{.85\columnwidth}{!}{  \begin{tabular}{|l|c|c|c|}
\hline
\multicolumn{1}{|c|}{\textbf{Model}} & \textbf{SECS} & \textbf{UTMOS} & \textbf{WER (\%)}\\
\hline
GT & $0.973\pm0.020$ & $2.814\pm1.011$ & $3.152$ \\
\hhline{|=|=|=|=|}
YourTTS & $0.940\pm0.041$ & $2.645\pm0.709$ & $5.313$\\
\hline
SpeechT5 & $\mathbf{0.969\pm0.013}$ & $2.064\pm0.706$ & $\mathbf{3.352}$\\
\hline
XTTS & $0.950\pm0.031$ & $\mathbf{2.746\pm0.857}$ & $8.030$\\
\hline
\end{tabular}}
\vspace{-0.15in}
\label{res:tts-human}
\end{center}
\end{table}

\begin{table}[t]  
\caption{Zero-shot TTS results for virtual speakers with 95\% CIs.}
\begin{center}
\resizebox{.85\columnwidth}{!}{ \begin{tabular}{|l|c|c|c|}
\hline
\multicolumn{1}{|c|}{\textbf{Model}} & \textbf{SECS} & \textbf{UTMOS} & \textbf{WER (\%)}\\
\hline
Virtual-GT& $0.977\pm0.010$ & $2.773\pm0.716$ & $2.740$ \\
\hhline{|=|=|=|=|}
YourTTS & $0.938\pm0.035$ & $\mathbf{2.755\pm0.627}$ & $5.298$\\
\hline
SpeechT5 & $\mathbf{0.961\pm0.011}$ & $2.221\pm0.684$ & $\mathbf{2.195}$\\
\hline
XTTS & $0.952\pm0.022$ & $2.696\pm0.708$ & $2.551$\\
\hline
\end{tabular}}
\vspace{-0.14in}
\label{res:tts-virtual}
\end{center}
\end{table}

\begin{table}[htp]
\caption{PSE results for the real-world speakers. M, S, and T in size indicate medium, small, and tiny sizes, respectively.}
\begin{center}
\resizebox{.85\columnwidth}{!}{  \begin{tabular}{|l|c|c|c|c|c|}
\hline
\multicolumn{1}{|c|}{\textbf{Model}} & \multicolumn{1}{c|}{\textbf{Size}} & \textbf{SDRI} & \textbf{SDR} & \textbf{eSTOI} & \textbf{PESQ} \\
\hline
\multirow{3}{*}{Generalist} & M & 9.495 & 9.997 & 0.708 & 1.487 \\
& S & 9.069 & 9.572 & 0.693 & 1.446 \\
& T & 8.267 & 8.770 & 0.670 & 1.382 \\
\hline
        \multirow{3}{*}{GT-6min}& M & 13.890 & 14.393 & 0.824 & 2.108 \\
& S & 12.991 & 13.494 & 0.799 & 1.952 \\
& T & 12.247 & 12.750 & 0.776 & 1.810 \\
        \hhline{|=|=|=|=|=|=|}
        \multirow{3}{*}{YourTTS-6min} & M & 12.110 & 12.613 & 0.786 & 1.950 \\
& S & 11.217 & 11.720 & 0.765 & 1.835 \\
& T & 10.521 & 11.024 & 0.746 & 1.721 \\
        \hline
        \multirow{3}{*}{YourTTS-30min}& M & 12.347 & 12.850 & 0.789 & 1.968 \\
& S & 11.427 & 11.931 & 0.770 & 1.857 \\
& T & 10.714 & 11.217 & 0.749 & 1.728 \\
        \hline
        \multirow{3}{*}{SpeechT5-6min} & M & 12.390 & 12.893 & 0.798 & 1.991 \\
& S & 11.479 & 11.982 & 0.774 & 1.847 \\
& T & 10.710 & 11.213 & 0.751 & 1.712 \\
        \hline
        \multirow{3}{*}{SpeechT5-30min}& M & \textbf{12.519} & \textbf{13.022} & \textbf{0.801} & 2.004 \\
& S & \textbf{11.626} & \textbf{12.129} & \textbf{0.778} & 1.862 \\
& T & \textbf{10.842} & \textbf{11.345} & \textbf{0.754} & 1.724 \\
\hline
        \multirow{3}{*}{XTTS-6min} & M & 12.302 & 12.805 & 0.795 & \textbf{2.013} \\
& S & 11.413 & 11.916 & 0.770 & 1.870 \\
& T & 10.514 & 11.017 & 0.743 & 1.735 \\
\hline
        \multirow{3}{*}{XTTS-30min}& M & 12.341 & 12.844 & 0.795 & \textbf{2.013} \\
& S & 11.453 & 11.956 & 0.771 & \textbf{1.877} \\
& T & 10.593 & 11.097 & 0.744 & \textbf{1.742} \\
        \hline
\end{tabular}}
\vspace{-0.14in}
\label{res:real}
\end{center}
\end{table}

\begin{table}[htp]
\caption{PSE results for the virtual speakers.}
\begin{center}
\resizebox{.85\columnwidth}{!}{  \begin{tabular}{|l|c|c|c|c|c|}
\hline
\multicolumn{1}{|c|}{\textbf{Model}} & \multicolumn{1}{c|}{\textbf{Size}} & \textbf{SDRI} & \textbf{SDR} & \textbf{eSTOI} & \textbf{PESQ} \\
\hline
\multirow{3}{*}{Generalist}& M & 10.291 & 10.749 & 0.814 & 1.467 \\
& S & 10.298 & 10.756 & 0.807 & 1.440 \\
& T & 9.455 & 9.912 & 0.788 & 1.368 \\
\hline
        \multirow{3}{*}{Virtual-GT-6min} & M & 14.907 & 15.365 & 0.890 & 1.986 \\
& S & 14.123 & 14.580 & 0.870 & 1.828 \\
& T & 13.396 & 13.854 & 0.850 & 1.679 \\
        \hhline{|=|=|=|=|=|=|}
        \multirow{3}{*}{YourTTS-6min}& M & 12.831 & 13.288 & 0.856 & 1.903 \\
& S & 12.567 & 13.025 & 0.841 & 1.776 \\
& T & 11.737 & 12.195 & 0.817 & 1.632 \\
        \hline
        \multirow{3}{*}{YourTTS-30min} & M & 12.960 & 13.418 & 0.859 & 1.908 \\
& S & 12.624 & 13.082 & 0.842 & 1.773 \\
& T & 11.768 & 12.226 & 0.818 & 1.625 \\
        \hline
        \multirow{3}{*}{SpeechT5-6min} & M & 13.682 & 14.140 & 0.873 & 1.915 \\
& S & 12.959 & 13.417 & 0.851 & 1.784 \\
& T & 12.188 & 12.646 & 0.832 & 1.636 \\
\hline
        \multirow{3}{*}{SpeechT5-30min}& M & \textbf{13.713} & \textbf{14.171} & \textbf{0.874} & 1.918 \\
& S & \textbf{12.979} & \textbf{13.437} & \textbf{0.852} & \textbf{1.784} \\
& T & \textbf{12.199} & \textbf{12.657} & \textbf{0.832} & \textbf{1.641} \\ \hline
        \multirow{3}{*}{XTTS-6min}& M & 13.288 & 13.746 & 0.866 & 1.910 \\
& S & 12.480 & 12.938 & 0.840 & 1.761 \\
& T & 11.824 & 12.282 & 0.821 & 1.627 \\ \hline
        \multirow{3}{*}{XTTS-30min} & M & 13.276 & 13.734 & 0.864 & \textbf{1.927} \\
& S & 12.479 & 12.937 & 0.839 & 1.758 \\
& T & 11.765 & 12.223 & 0.818 & 1.626 \\ 
        \hline 
\end{tabular}}
\vspace{-0.15in}
\label{res:virtual}
\end{center}
\end{table}

\subsection{Results}
\noindent\textbf{Basic TTS performance}: 
The performance of zero-shot TTS models for real-world and virtual speakers is detailed in Table \ref{res:tts-human} and Table \ref{res:tts-virtual}, respectively. SpeechT5 achieves the best speaker similarity (SECS) and intelligibility (WER) scores, despite having the lowest perceptual quality (UTMOS) scores in both cases. For real-world speakers, XTTS achieves the highest UTMOS score but the worst WER score, while for virtual speakers, it ranks second in both the UTMOS score and WER score. We assume that the diversity of accents and emotions among the virtual speakers, along with some artificial noise already present in the reference speech, influenced these performance outcomes. Overall, each TTS system has its own unique property, which can be measured in different ways. Next, we examine how these properties affect the downstream task performance when used as a data augmentation method.

\noindent\textbf{PSE performance}: The PSE results for real-world and virtual speakers are detailed in Table \ref{res:real} and Table \ref{res:virtual}, respectively. Across all model sizes, the generalist models performed the worst on every evaluation metric. This implies that even PSE models built using the lowest-performing zero-shot TTS systems achieved significantly better performance than the generalist SE model. This demonstrates the effectiveness of personalized data augmentation using external generative models. For all sizes, the \textbf{GT-6min} model achieved the highest scores across all metrics, outperforming all \textbf{30min} models. This suggests that the data quality is crucial for PSE performance; even with a larger dataset of lower quality, performances were inferior to those achieved with a smaller amount of high-quality data. Compared to \cite{minje_pse_2}, where some TTS models did not introduce PSE improvement, this time, the adaptation to the noise sources could have contributed to better PSE performance. 

When comparing the \textbf{6min} and \textbf{30min} models fine-tuned with augmented data from various zero-shot TTS models, we observed that in most cases, PSE performance improved as the amount of the augmented data increased, although the improvement is marginal. For both real-world and virtual speakers, the \textbf{SpeechT5-30min} model achieved the best performance in SDRI, SDR, and eSTOI. Given the high speaker similarity of the \textbf{SpeechT5} model, we believe that speaker similarity is an important factor in building an effective PSE model.
The \textbf{XTTS-30min} model achieved the highest PESQ scores for real-world speakers and for medium-sized PSE model for virtual speakers, while the \textbf{SpeechT5-30min} model performed best for small and tiny-sized models for virtual speakers. Since PESQ focuses on perpetual quality, the \textbf{XTTS} model's high perceptual quality likely contributed to its strong PESQ performance.

\section{Discussion and Future Work}
In our baseline experiments, we demonstrated the potential of zero-shot TTS models for data augmentation in PSE applications. We also highlighted the importance of adaptation data quality for PSE model performance. Speaker similarity and intelligibility emerged as the most relevant factors, with perceptual quality also influencing PSE outcomes. However, as the number of zero-shot TTS models in our baseline experiments was limited, we anticipate that this challenge will enable a more in-depth exploration of the relationship between TTS model performance and PSE outcomes through a broader variety of zero-shot TTS systems. Data augmentation stands to benefit significantly from advances in generative AI, though its application requires careful consideration due to the complex nature of synthetic data usability. {We also explored virtual speakers as a privacy-preserving alternative. A possible application is to build a PSE model that reflects the target speaker’s characteristics using virtual speakers, thereby addressing privacy concerns associated with collecting target speaker data.} In the future, the organizers plan to expand this challenge to additional downstream tasks.

\bibliographystyle{IEEEtran}
\bibliography{test}

\begin{thebibliography}{10}
\providecommand{\url}[1]{#1}
\csname url@samestyle\endcsname
\providecommand{\newblock}{\relax}
\providecommand{\bibinfo}[2]{#2}
\providecommand{\BIBentrySTDinterwordspacing}{\spaceskip=0pt\relax}
\providecommand{\BIBentryALTinterwordstretchfactor}{4}
\providecommand{\BIBentryALTinterwordspacing}{\spaceskip=\fontdimen2\font plus
\BIBentryALTinterwordstretchfactor\fontdimen3\font minus \fontdimen4\font\relax}
\providecommand{\BIBforeignlanguage}[2]{{%
\expandafter\ifx\csname l@#1\endcsname\relax
\typeout{** WARNING: IEEEtran.bst: No hyphenation pattern has been}%
\typeout{** loaded for the language `#1'. Using the pattern for}%
\typeout{** the default language instead.}%
\else
\language=\csname l@#1\endcsname
\fi
#2}}
\providecommand{\BIBdecl}{\relax}
\BIBdecl

\bibitem{whisper}
A.~Radford, J.~W. Kim, T.~Xu, G.~Brockman, C.~Mcleavey \emph{et~al.}, ``Robust speech recognition via large-scale weak supervision,'' in \emph{Proc. Int. Conf. on Machine Learning (ICLR)}, 2023.

\bibitem{vits}
J.~Kim, J.~Kong, and J.~Son, ``Conditional variational autoencoder with adversarial learning for end-to-end text-to-speech,'' in \emph{Proc. Int. Conf. on Machine Learning (ICLR)}, 2021.

\bibitem{yourtts}
E.~Casanova, J.~Weber, C.~D. Shulby, A.~C. Junior, E.~G{\"o}lge \emph{et~al.}, ``{Y}our{TTS}: Towards zero-shot multi-speaker {TTS} and zero-shot voice conversion for everyone,'' in \emph{Proc. Int. Conf. on Machine Learning (ICML)}, 2022.

\bibitem{xtts}
E.~Casanova, K.~Davis, E.~Gölge, G.~Göknar, I.~Gulea \emph{et~al.}, ``{XTTS}: A massively multilingual zero-shot text-to-speech model,'' in \emph{Proc. Interspeech}, 2024.

\bibitem{valle}
\BIBentryALTinterwordspacing
C.~Wang, S.~Chen, Y.~Wu, Z.~Zhang, L.~Zhou, S.~Liu \emph{et~al.}, ``Neural codec language models are zero-shot text to speech synthesizers,'' \emph{CoRR}, vol. abs/2301.02111, 2023. [Online]. Available: \url{https://doi.org/10.48550/arXiv.2301.02111}
\BIBentrySTDinterwordspacing

\bibitem{se_1}
S.-W. Fu, C.~Yu, T.-A. Hsieh, P.~Plantinga, M.~Ravanelli \emph{et~al.}, ``{MetricGAN+}: An improved version of {MetricGAN} for speech enhancement,'' in \emph{Proc. Interspeech}, 2021.

\bibitem{se_2}
H.~J. Park, B.~H. Kang, W.~Shin, J.~S. Kim, and S.~W. Han, ``{MANNER}: Multi-view attention netwfork for noise erasure,'' in \emph{Proc. IEEE Int. Conf. on Acoustics, Speech and Signal Processing (ICASSP)}, 2022.

\bibitem{apple_pasr}
K.~Yang, T.-Y. Hu, J.-H.~R. Chang, H.~Swetha~Koppula, and O.~Tuzel, ``Text is all you need: Personalizing {ASR} models using controllable speech synthesis,'' in \emph{Proc. IEEE Int. Conf. on Acoustics, Speech and Signal Processing (ICASSP)}, 2023.

\bibitem{minje_pse_2}
A.~Kuznetsova, A.~Sivaraman, and M.~Kim, ``The potential of neural speech synthesis-based data augmentation for personalized speech enhancement,'' in \emph{Proc. IEEE Int. Conf. on Acoustics, Speech and Signal Processing (ICASSP)}, 2023.

\bibitem{image_dataaug_survey}
C.~Shorten and T.~M. Khoshgoftaar, ``A survey on image data augmentation for deep learning,'' \emph{Journal of Big Data}, vol.~6, p.~60, 2019.

\bibitem{image_da_diffusion}
B.~Trabucco, K.~Doherty, M.~A. Gurinas, and R.~Salakhutdinov, ``Effective data augmentation with diffusion models,'' in \emph{Proc. Int. Conf. on Learning Representations (ICLR)}, 2024.

\bibitem{spechaug}
D.~S. Park, W.~Chan, Y.~Zhang, C.-C. Chiu, B.~Zoph \emph{et~al.}, ``{SpecAugment}: A simple data augmentation method for automatic speech recognition,'' in \emph{Proc. Interspeech}, 2019.

\bibitem{latent_filling}
J.-S. Bae, J.~Y. Lee, J.-H. Lee, S.~Mun, T.~Kang \emph{et~al.}, ``Latent filling: Latent space data augmentation for zero-shot speech synthesis,'' in \emph{Proc. IEEE Int. Conf. on Acoustics, Speech and Signal Processing (ICASSP)}, 2024.

\bibitem{tts_for_DA}
Q.~Chen, Z.~Ma, T.~Liu, X.~Tan, Q.~Lu \emph{et~al.}, ``Improving few-shot learning for talking face system with {TTS} data augmentation,'' in \emph{Proc. IEEE Int. Conf. on Acoustics, Speech and Signal Processing (ICASSP)}, 2023.

\bibitem{tts_keyword_da}
\BIBentryALTinterwordspacing
H.~J. Park, D.~Agarwal, N.~Chen, R.~Sun, K.~Partridge \emph{et~al.}, ``Utilizing {TTS} synthesized data for efficient development of keyword spotting model,'' \emph{CoRR}, vol. abs/2407.18879, 2024. [Online]. Available: \url{https://doi.org/10.48550/arXiv.2407.18879}
\BIBentrySTDinterwordspacing

\bibitem{ZSTTS_DA_for_ASR}
F.~Nespoli, D.~Barreda, and P.~A. Naylor, ``Zero shot text to speech augmentation for automatic speech recognition on low-resource accented speech corpora,'' in \emph{57th Asilomar Conference on Signals, Systems, and Computers}, 2023.

\bibitem{libritts}
H.~Zen, R.~Clark, R.~J. Weiss, V.~Dang, Y.~Jia \emph{et~al.}, ``{LibriTTS}: A corpus derived from librispeech for text-to-speech,'' in \emph{Proc. Interspeech}, 2019.

\bibitem{musan}
\BIBentryALTinterwordspacing
D.~Snyder, G.~Chen, and D.~Povey, ``{MUSAN:} {A} music, speech, and noise corpus,'' \emph{CoRR}, vol. abs/1510.08484, 2015. [Online]. Available: \url{http://arxiv.org/abs/1510.08484}
\BIBentrySTDinterwordspacing

\bibitem{speechbrain}
\BIBentryALTinterwordspacing
M.~Ravanelli, T.~Parcollet, P.~Plantinga, A.~Rouhe, S.~Cornell \emph{et~al.}, ``{SpeechBrain}: {A} general-purpose speech toolkit,'' \emph{CoRR}, vol. abs/2106.04624, 2021. [Online]. Available: \url{https://arxiv.org/abs/2106.04624}
\BIBentrySTDinterwordspacing

\bibitem{mosnet}
Y.~Choi, Y.~Jung, and H.~Kim, ``Deep {MOS} predictor for synthetic speech using cluster-based modeling,'' in \emph{Proc. Interspeech}, 2020.

\bibitem{mosnet2}
G.~Mittag, B.~Naderi, A.~Chehadi, and S.~Moller, ``{NISQA}: A deep {CNN}-self-attention model for multidimensional speech quality prediction with crowdsourced datasets,'' in \emph{Proc. Interspeech}, 2021.

\bibitem{svc_utmos1}
W.-C. Huang, L.~P. Violeta, S.~Liu, J.~Shi, and T.~Toda, ``The singing voice conversion challenge 2023,'' in \emph{Proc. IEEE Automatic Speech Recognition and Understanding Workshop (ASRU)}, 2023.

\bibitem{naturalspeech3_utmos2}
Z.~Ju, Y.~Wang, K.~Shen, X.~Tan, D.~Xin \emph{et~al.}, ``{NaturalSpeech 3}: Zero-shot speech synthesis with factorized codec and diffusion models,'' in \emph{Proc. Int. Conf. on Machine Learning (ICML)}, 2024.

\bibitem{utmosv2}
\BIBentryALTinterwordspacing
K.~Baba, W.~Nakata, Y.~Saito, and H.~Saruwatari, ``The {T05} system for the {VoiceMOS} challenge 2024: Transfer learning from deep image classifier to naturalness {MOS} prediction of high-quality synthetic speech,'' \emph{CoRR}, vol. abs/2409.09305, 2024. [Online]. Available: \url{https://doi.org/10.48550/arXiv.2409.09305}
\BIBentrySTDinterwordspacing

\bibitem{voicemos_challenge}
\BIBentryALTinterwordspacing
W.~Huang, S.~Fu, E.~Cooper, R.~E. Zezario, T.~Toda, H.~Wang, J.~Yamagishi, and Y.~Tsao, ``The {VoiceMOS} challenge 2024: Beyond speech quality prediction,'' \emph{CoRR}, vol. abs/2409.07001, 2024. [Online]. Available: \url{https://doi.org/10.48550/arXiv.2409.07001}
\BIBentrySTDinterwordspacing

\bibitem{sdr}
E.~Vincent, R.~Gribonval, and C.~Fevotte, ``Performance measurement in blind audio source separation,'' \emph{IEEE Transactions on Audio, Speech, and Language Processing}, vol.~14, no.~4, pp. 1462--1469, 2006.

\bibitem{estoi}
C.~H. Taal, R.~C. Hendriks, R.~Heusdens, and J.~Jensen, ``A short-time objective intelligibility measure for time-frequency weighted noisy speech,'' in \emph{Proc. IEEE Int. Conf. on Acoustics, Speech and Signal Processing (ICASSP)}, 2010.

\bibitem{pesq}
A.~Rix, J.~Beerends, M.~Hollier, and A.~Hekstra, ``Perceptual evaluation of speech quality {(PESQ)}-a new method for speech quality assessment of telephone networks and codecs,'' in \emph{Proc. IEEE Int. Conf. on Acoustics, Speech, and Signal Processing. (ICASSP)}, 2001.

\bibitem{speecht5}
J.~Ao, R.~Wang, L.~Zhou, C.~Wang, S.~Ren \emph{et~al.}, ``{S}peech{T}5: Unified-modal encoder-decoder pre-training for spoken language processing,'' in \emph{Proc. of the 60th Annual Meeting of the Association for Computational Linguistics}, 2022.

\bibitem{tortoise}
\BIBentryALTinterwordspacing
J.~Betker, ``Better speech synthesis through scaling,'' \emph{CoRR}, vol. abs/2305.07243, 2023. [Online]. Available: \url{https://doi.org/10.48550/arXiv.2305.07243}
\BIBentrySTDinterwordspacing

\bibitem{convtasnet}
Y.~Luo and N.~Mesgarani, ``{Conv-TasNet}: Surpassing ideal time–frequency magnitude masking for speech separation,'' \emph{IEEE/ACM Transactions on Audio, Speech, and Language Processing}, vol.~27, no.~8, pp. 1256--1266, 2019.

\bibitem{minje_pse_1}
A.~Sivaraman and M.~Kim, ``Efficient personalized speech enhancement through self-supervised learning,'' \emph{IEEE Journal of Selected Topics in Signal Processing}, vol.~16, no.~6, pp. 1342--1356, 2022.

\bibitem{librispeech}
V.~Panayotov, G.~Chen, D.~Povey, and S.~Khudanpur, ``Librispeech: An {ASR} corpus based on public domain audio books,'' in \emph{Proc. IEEE Int. Conf. on Acoustics, Speech and Signal Processing (ICASSP)}, 2015.

\bibitem{fsdk50}
E.~Fonseca, X.~Favory, J.~Pons, F.~Font, and X.~Serra, ``{FSD50K}: An open dataset of human-labeled sound events,'' \emph{IEEE/ACM Transactions on Audio, Speech, and Language Processing}, vol.~30, pp. 829--852, 2022.

\bibitem{adam}
\BIBentryALTinterwordspacing
D.~P. Kingma and J.~Ba, ``Adam: {A} method for stochastic optimization,'' in \emph{Proc. Int. Conf. on Learning Representations (ICLR)}, 2015. [Online]. Available: \url{http://arxiv.org/abs/1412.6980}
\BIBentrySTDinterwordspacing

\end{thebibliography}

\end{document}